\documentclass[aps,prd,preprint,floatfix,nofootinbib,longbibliography,a4paper ]{revtex4-1}
\pdfoutput=1
\usepackage{color}
\usepackage{graphicx}
\usepackage{textcomp}
\usepackage{subfigure}
\usepackage{feynmp}
\usepackage{amsmath,amssymb,array}
\usepackage{enumerate}
\usepackage{hhline}
\newcommand{\mathsym}[1]{{}}

\def\10{$SO(10)$}

\newcommand{\ba}{\begin{array}}
\newcommand{\ea}{\end{array}}
\newcommand{\be}{\begin{equation}}
\newcommand{\ee}{\end{equation}}
\newcommand{\beqa}{\begin{eqnarray}}
\newcommand{\eeqa}{\end{eqnarray}}
\def\321{$SU(3)\times SU(2)\times U(1)$}
\def\b126{$\overline{126}$}
\def\lmt{$L_\mu-L_\tau$~}
\def\mt{$\mu$-$\tau$~}

\def\vev#1{\left\langle #1\right\rangle}%

\usepackage{accents}

\begin{document}

\vspace*{0.5cm}
\title{Generalised $\mu$-$\tau$ symmetries and calculable gauge kinetic and mass mixing in $U(1)_{L_\mu - L_\tau}$ models}

\author{Anjan S. Joshipura}
\email{anjan@prl.res.in}
\author{Namit Mahajan}
\email{nmahajan@prl.res.in}
\author{Ketan M. Patel}
\email{kmpatel@prl.res.in}
\affiliation{Physical Research Laboratory, Navarangpura, Ahmedabad-380 009, India}

\begin{abstract}
Extensions of the standard model with a $U(1)$ gauge symmetry contain gauge invariant kinetic mixing, $\sin\chi$,
and gauge non-invariant mass mixing, $\delta M^2$, between the hypercharge and the new gauge boson $Z'$. 
These represent a priori incalculable but phenomenologically important parameters of the theory. They become calculable if
there exist spontaneously or softly broken symmetries which forbid them at tree level but allow their generation at the loop level. 
We discuss various symmetries falling in this category in the context of the gauged $L_\mu-L_\tau$ models and their interplay with lepton mixing. 
It is shown that one gets phenomenologically inconsistent lepton mixing parameters if these symmetries are exact. 
Spontaneous breaking of these symmetries can lead to consistent lepton mixing and also generates finite and calculable values of 
these parameters at one or two loop order depending on the underlying symmetry. 
We calculate these parameters in two specific cases: (i) the standard seesaw model with $\mu$-$\tau$  symmetry broken by the masses of 
the right-handed neutrinos and (ii) in a model containing a pair of vectorlike charged leptons which break $\mu$-$\tau$ symmetry.
In case (i), the right-handed neutrinos are the only source of gauge mixing. The kinetic mixing parameters are suppressed and vanish 
if the right-handed neutrinos decouple from the theory. In contrast, there exists a finite gauge mixing in case 
(ii) which survives even when the masses of vectorlike leptons are taken to infinity, exhibiting non-decoupling behaviour. 
The seesaw model discussed here represents a complete framework with practically no kinetic mixing and hence can survive a 
large number of experimental probes used to rule out specific ranges in the coupling $g^\prime$ and mass $M_{Z^\prime}$.
The model can generate non-universality in tau decays, which can be tested in future experiments. 
\end{abstract}

\maketitle

\section{Introduction}
\label{sec:intro}
The observed deviations from the Standard Model (SM) predictions in the anomalous magnetic moment of muon, $B$ meson decays, 
and the need to explain the origin of the dark matter in the universe has led to various extensions of the SM gauge symmetry. 
One of the simplest extensions corresponds to the addition of a $U(1)$ gauge group and the most economical among them are the ones
which do not require extension in the fermionic content of the SM. Three such possibilities corresponding to differences in
the leptonic flavour indices $L_\alpha-L_\beta$ have been identified long ago \cite{Foot:1990mn,He:1990pn,He:1991qd}.
Most conspicuous of these three choices is the $U(1)$ gauge group corresponding to \lmt. In the simplest form, the gauge
boson of \lmt does not couple to quarks and the first generation leptons, thereby avoiding many constraints coming from these sectors. 
Phenomenology of \lmt gauge symmetry has been extensively discussed in a number of papers (see for incomplete list of references \cite{Foot:1994vd,Baek:2001kca,Ma:2001md,Baek:2001kca,Joshipura:2003jh,Adhikary:2006rf,Chun:2007vh,Baek:2008nz,Heeck:2010pg,Heeck:2010pg,Heeck:2011wj,Harigaya:2013twa,Das:2013jca,Chatterjee:2015gta,Altmannshofer:2014cfa,Altmannshofer:2016brv,Altmannshofer:2016jzy,Biswas:2016yjr,Ibe:2016dir,Patra:2016shz,Araki:2017wyg,Chen:2017gvf,Chen:2017usq,Nomura:2018vfz,Arcadi:2018tly,Banerjee:2018eaf,Nomura:2018cle,Singirala:2018mio,Banerjee:2018mnw,Biswas:2019twf,Escudero:2019gzq,Poddar:2019wvu,Heeck:2018nzc}) in various contexts.

The additional $U(1)$ symmetry may be broken at a scale smaller than the electroweak scale in which case the new interactions are 
constrained by a variety of low energy processes. The new gauge boson having mass in the range $100$-$400$ MeV is advocated 
(see, for example, \cite{Gninenko:2018tlp}) as an interesting possibility in case of the \lmt symmetry. 
Such a gauge boson is consistent with various constraints from the laboratory experiments and may also explain the possible 
discrepancy between the measured value of the muon $(g-2)$ and that predicted in the SM \cite{Baek:2001kca}. Alternatively, if $U(1)_{L_\mu-L_\tau}$ is broken at a scale significantly larger than the electroweak scale, then all the effects associated with the new gauge boson would be suppressed by its mass. Such effects would appear as non-renomalizable operators in the effective theory below the \lmt breaking scale. An exception to this, in case of all the $U(1)$ gauge groups, is a dimension four operator allowed by gauge symmetries. It is conveniently parameterized as \cite{Holdom:1985ag,delAguila:1988jz,Babu:1996vt}
\be 
\label{km}-\frac{\sin\chi}{2} F_{\mu\nu}^B F^{\mu\nu}_{Z'}\,
\ee
where $F_{\mu\nu}^B, F^{\mu\nu}_{Z'}$ are the field strength tensors for the hypercharge and the $Z'$ gauge bosons respectively. An associated operator which can result after breaking of hypercharge and $U(1)$ symmetry is the mass mixing among two gauge bosons $B$ and $Z'$:
\be \label{mass}
\delta M^2\, B^\mu Z^{\prime}_ \mu ~.
\ee
These two parameters provide a window into new gauge symmetry if it is broken at a very large scale. 

The Kinetic Mixing (KM) parameters, $\sin\chi$ and $\delta M^2$, are arbitrary a priori but can be
constrained from various experiments. The main effect of these parameters is to
mix the additional gauge boson with $Z$ thereby coupling quarks and
electron to $Z'$ and changing the existing couplings of $Z$ to fermions.
This leads to observable effects in precision electroweak tests, atomic
parity violation \cite{Babu:1996vt}, neutrino trident production
\cite{Altmannshofer:2014pba}, the low energy elastic scattering of the
solar neutrino with electrons observed in Borexino \cite{Kaneta:2016uyt} and
coherent elastic $\nu N$ scattering \cite{Abdullah:2018ykz}. One obtains
significant bounds on $\sin\chi$ from these processes. 
These are model dependent. A summary of various constraints can be found, for example, in \cite{Bauer:2018onh}. 
In models with a diagonal charged lepton mass matrix, one obtains
\cite{Bauer:2018onh,Gninenko:2018tlp}: $\sin\chi\sim 10^{-5}-10^{-6}$ for
$M_{Z^\prime} \sim 100-200$  MeV.  One can obtain some meaningful
predictions for $\sin\chi$ and $\delta M^2$ by invoking discrete symmetry
which forbids them at the tree level. If this symmetry remains unbroken, then
the \lmt symmetry broken at a very high scale would remain completely
hidden. On the other hand, the spontaneous breaking of such discrete
symmetry would lead to calculable values for the said parameters. One possible symmetry forbidding Eqs. (\ref{km},\ref{mass}) was considered first in
\cite{Foot:1994vd}. This corresponds to interchanging $\mu$ and $\tau$
degrees accompanied by a change in sign of $Z'$. One could consider various
generalizations of this symmetry any of which can be used to forbid
interactions in Eqs. (\ref{km},\ref{mass}) at the tree level. One of the aims of this paper is to discuss possible classes of symmetries which forbid
Eqs. (\ref{km},{\ref{mass}) and their implications for the leptonic mixing.
The symmetries invoked to forbid KM parameters directly
influence the neutrino mixing pattern since they also constrain the leptonic
Yukawa couplings. We show that none of the symmetries used here to forbid
the KM parameters at tree level can remain unbroken if one is to obtain consistent mixing angles and CP violation in the leptonic sector. The breaking of
these symmetries then generates the KM parameters at the 1-loop or at the 2-loop level
as we discuss. 
 
KM parameter $\sin\chi$ gets generated at the 1-loop level by the charged
leptons in the  standard scenario adopted in many works
\cite{Holdom:1985ag,Gninenko:2018tlp,Kamada:2015era,Ibe:2016dir,
Araki:2017wyg}.
This happens however for a specific case in which the charged lepton mass
matrix is \lmt invariant and hence diagonal and break \mt symmetry. The neutrino mass matrix in
this case cannot also be invariant under the \mt symmetry if it is to reproduce
the observed leptonic mixing angles. One gets an additional contribution to
KM from this mass matrix. The neutrino contribution to $\sin\chi$ is
suppressed by the neutrino masses, i.e. $\sin\chi\sim {\cal O}(\frac{m_\nu^2}{q^2})$, but the contribution to $\delta M^2$ contains a divergent piece if the charged leptons and three light Majorana neutrinos
are the only sources which generate Eqs. (\ref{km},\ref{mass}). This makes the resulting $Z$-$Z'$ mixing incalculable. The 
divergent contribution can be removed only in a complete theory with
spontaneously or softly broken \mt symmetry. The other aim of the paper is
to provide explicit models within which all the infinities which would arise through ad-hoc breaking of \mt symmetry vanish. 
We present two specific examples, one which leads to
unobservablly  small and the other with a fairly large value of
$\sin\chi$. The first example is the standard seesaw model in which the
source of \mt breaking is confined in the Right-Handed (RH) neutrino mass matrix. The
standard contribution from the charged leptons considered in the literature
actually vanish in this case. We present detailed calculations showing that
one gets finite  $\sin\chi$ and $\delta M^2$ at the 1-loop from the neutrino
sector. Both these parameters are suppressed in the model and vanish in the limit of very large RH neutrino masses.  However, $\delta M^2$ can get finite non-decoupling contribution, independent of the RH neutrino masses, in the presence of \mt symmetry breaking in the Dirac neutrino mass matrix. The second example we discuss corresponds to adding the charged vectorlike fermions to
the SM. Their mass terms, allowed by gauge symmetry, provide the only source of the \mt  (or other analogous symmetry) breaking and are responsible for generation of finite contribution to mixing parameters. This model shows
the non-decoupling effects and contains a direct contribution from the
vectorlike fermions, which does not vanish when vectorlike masses are taken
to infinity. This contribution thus could be large.

The paper is organized as follows. We discuss symmetries which lead to
vanishing KM in section \ref{sec:symmetries} and their
consequences on the leptonic mixing in section \ref{sec:leptonmixing}.
Evaluation of KM at 1-loop in a general framework is given in
section \ref{sec:general} and two specific examples are discussed in details
in section \ref{sec:model}. We summarize our results in \ref{sec:summary}
and give a short discussion of the already existing literature  of the
phenomenology of the type of models discussed here.

\section{Symmetries for vanishing kinetic mixing in \lmt model}
\label{sec:symmetries}
The minimal version of the \lmt model is obtained by assigning equal and
opposite $U(1)$ charges to the leptonic doublets $L_\mu^\prime$ and
$L_\tau^\prime$, where $L^\prime_\alpha = (\nu_{\alpha L}^\prime, l_{\alpha
L}^\prime)^T~,\alpha=e,\mu,\tau$. The right handed charged leptons $l_{\mu
R}^\prime$ and $l_{\tau R}^\prime$ carry the same \lmt charges as
$L_\mu^\prime$ and $L_\tau^\prime$, respectively. Rest of the SM fermions
are neutral under the additional $U(1)$. The neutral current interactions of
leptons are then given by
\be \label{nc}
-{\cal L}_{\rm NC}=g_Y B_\mu\left(-\frac{1}{2}\overline{L}_\alpha^\prime
\gamma^\mu L_\alpha^\prime - \overline{l}_{\alpha R}^\prime \gamma^\mu
l_{\alpha R}^\prime\right)
+g^\prime Z^\prime_\mu\left(\overline{l}_\mu^\prime \gamma^\mu l_\mu^\prime
-\overline{l}_\tau^\prime \gamma^\mu l_\tau^\prime +
\overline{\nu}^\prime_{\mu L} \gamma^\mu \nu^\prime_{\mu
L}-\overline{\nu}^\prime_{\tau L}\gamma^\mu \nu^\prime_{\tau L}\right)\,,\ee
 ${\cal L}_{\rm NC}$ is  invariant under the following transformations
\cite{Foot:1994vd}:
\be \label{mutau}
L^\prime _\mu\leftrightarrow L^\prime_\tau,~~l_{\mu R}^\prime
\leftrightarrow l_{\tau R}^\prime,~~ B_\mu\rightarrow B_\mu,~~
Z'_\mu\rightarrow - Z'_\mu\,.\ee
This symmetry acts as the standard \mt interchange symmetry on the leptons.
It also changes the sign of the new gauge boson in addition. The \mt
interchanges symmetry forbids the kinetic and mass mixing terms of Eqs.
(\ref{km},\ref{mass}) at tree level. If this symmetry is also respected by
the Higgs sector  and the Yukawa couplings of leptons then the entire
Lagrangian is invariant under it, and the KM remains absent to
all orders. If this is not the case, the KM will get generated
at the loop level. It is easy to derive conditions under which the leptonic
contribution to KM remains zero at 1-loop.

We collectively represent $l_{\alpha L}^\prime$, $l_{\alpha R}^\prime$,
$\nu_{\alpha L}^\prime$ as $f_\alpha^\prime$. The mixing matrix $U_f$
connecting the mass eigenstates $f_i$ of fermions to the weak eigenstates
$f_\alpha^\prime$ is defined as
\be \label{uf}
f^\prime_\alpha =(U_f)_{\alpha i } f_i\,.\ee
In the mass basis, the couplings to $Z^\prime$ boson given in Eq. (\ref{nc})
change to
\be \label{zpmass}
g' Z^\prime_\mu\, F^f_{ij}\, \overline{f}_i\gamma^\mu f_j\,, \ee
with
\be \label{falpha}
F^f_{ij}=
(U_f)^*_{\mu i}(U_f)_{\mu j}-(U_f)^*_{\tau i}(U_f)_{\tau j } ~.\ee 
The diagonal couplings, $F_{ii}^f$, vanish if
\be \label{mutaurgen}
|(U_f)_{\mu i}|^2=|(U_f)_{\tau i}|^2~.\ee
This equation, termed as the \mt reflection symmetry,  has useful
phenomenological consequences
\cite{Harrison:2002et,Grimus:2003yn,Joshipura:2015dsa} when
applied to the leptonic mixing matrix $U_{\rm PMNS}$. It has implication for
the KM as well. Since $B$ has only flavour diagonal couplings, and $Z'$ has
only off-diagonal couplings when Eq. (\ref{mutaurgen}) is satisfied, the
fermion loop connecting them in vacuum polarization diagram is absent and
the KM cannot arise at the 1-loop level. This however requires that Eq.
(\ref{mutaurgen}) holds individually for all the fermion mixing matrices
$U_{l_L}$, $U_{l_R}$ and $U_{\nu_L}$. Eq. (\ref{mutaurgen}) in this case
represents sufficient conditions for vanishing of the leptonic contribution
to  KM parameters  at the 1-loop level.

The \mt interchange symmetry is a special case leading to Eq.
(\ref{mutaurgen}). If the Yukawa couplings of leptons and the Higgs sector 
respect this symmetry then the Majorana neutrino and the charged lepton mass matrices
$M_\nu$ and $M_l$ respectively satisfy \cite{Gupta:2013it}
\be \label{mutaumass}
S^T M_\nu S=M_\nu\,,~~ S^\dagger M_l M_l^\dagger
S=M_lM_l^\dagger\,,~S^\dagger M_l^\dagger M_l S=M_l^\dagger M_l~,\ee
where 
\be \label{s}
S=\left(\ba{ccc}
1&0&0\\
0&0&1\\
0&1&0\\ \ea
\right)\ee
is the \mt interchange matrix. These equations lead to the corresponding
diagonalizing matrices as
\be \label{umutau}
U_{l_L} = U^{\mu \tau}(\theta_{L})Q_L\,,~~U_{l_R} = U^{\mu
\tau}(\theta_{R})Q_R\,,~~U_{\nu_L} = U^{\mu \tau}(\theta_\nu)Q_\nu\,, \ee
where $Q_L$, $Q_R$, $Q_\nu$ are diagonal phase matrices and
\be \label{umt}
U^{\mu\tau}(\theta)= \left( \ba{ccc}
\cos \theta &-\sin \theta & 0\\
\frac{1}{\sqrt{2}} \sin\theta & \frac{1}{\sqrt{2}} \cos \theta
&-\frac{1}{\sqrt{2}}\\
\frac{1}{\sqrt{2}} \sin\theta & \frac{1}{\sqrt{2}} \cos \theta
&\frac{1}{\sqrt{2}}\\ \ea
\right)\,.\ee
Each of the unitary matrices in Eq. (\ref{umutau}) satisfy the conditions depicted in Eq.
(\ref{mutaurgen}). Both the kinetic and mass mixing vanish in this case to all
orders as long as symmetry in  Eq. (\ref{mutau}) remains unbroken.

Eq. (\ref{mutaurgen}) provides clue to other possible symmetries which can
be used to forbid KM. One such well-studied example \cite{Grimus:2003yn}
corresponds to imposing the following symmetry on the leptonic fields
\be \label{mutaur1}
L_e^\prime \leftrightarrow L_e^{\prime\, CP},~~e_R^\prime \leftrightarrow
e_R^{\prime\, CP},~~L_\mu^\prime\leftrightarrow L_\tau^{\prime CP},~~l_{\mu
R}^\prime \leftrightarrow
l_{\tau R}^{\prime CP},\ee
where $f^{\prime CP} = \gamma^0 C \overline{f}^{\prime T}$. If this symmetry
is respected by the vacuum and Yukawa interactions, then the leptonic mass
matrices satisfy
\be \label{mutaur}
S^TM_\nu S=M_\nu^*,~~S^\dagger M_l M_l^\dagger
S=(M_lM_l^\dagger)^*,~~S^\dagger M_l^\dagger M_l S=(M_l^\dagger M_l)^*.\ee
The first of the above has been extensively studied in the diagonal basis of
the charged leptons \cite{Harrison:2002et,Grimus:2003yn}. Forbidding kinetic
mixing
would require that the entire Eq. (\ref{mutaur}) be satisfied
simultaneously. Above conditions imply \cite{Harrison:2002et,Grimus:2003yn}
that
the mixing matrices $U_{l_{L,R}}$, $U_{\nu}$ have the  form:
\be \label{umutaur}
U_{l_L} = U_{l_L} ^{\rm HS}Q_L\,,~~U_{l_R} = U_{l_R}^{\rm
HS}Q_R\,,~~U_{\nu_L} = U_{\nu_L}^{\rm HS} X_\nu\,, \ee
where  $Q_{L,R}$ are diagonal matrices of unphysical phases and  $X_\nu$ is
a diagonal matrix with $X_\nu^2=1$. The matrices $U_{l_L}^{\rm HS}$,
$U_{l_R}^{\rm HS}$ and $U_{\nu_L}^{\rm HS}$  possess the Harrison-Scott form
\cite{Harrison:2002et}
\be \label{hs}
U^{\rm HS}=\left ( \ba{ccc}
x_1&x_2&x_3\\
z_1&z_2&z_3\\
z_1^*&z_2^*&z_3^*\\ \ea \right)~.\ee
with $x_{1,2,3}$ real. The above form of $U_{l_{L,R}}$, $U_{\nu_L}$
satisfies Eq. (\ref{mutaurgen}) and the KM cannot arise at the 1-loop
level. Eq. (\ref{mutaur}) is more general and can forbid the KM to all
orders. This follows from the fact that the neutral current interactions in
Eq. (\ref{nc}) are invariant if the leptonic symmetry, Eq. (\ref{mutaur1}),
is supplemented with the following transformation on the  gauge bosons $B$,
$Z'$:
\be \label{mutaur2}
 Z'_\mu\rightarrow Z'^\mu,~~ B_\mu\rightarrow - B^\mu\,.\ee
Action of this symmetry on $B_\mu$ corresponds to the standard CP
transformation and thus CP invariance of the gauge interactions assures the
above mentioned symmetry for the hypercharge current of fermions and scalars. But the
$Z'$ needs to be transformed in the opposite manner compared to $B$ to make the
corresponding term in Eq. (\ref{nc}) invariant under this generalised CP.
This ensures that the KM parameters remain zero to all orders as
long as Eq. (\ref{mutaur})  holds and the Higgs sector also respects
appropriately defined \mt reflection symmetry.

\section{Vanishing kinetic mixing and leptonic mixing}
\label{sec:leptonmixing}
The two examples of symmetries discussed in the previous section which forbid the
KM in \lmt model have implications on the leptonic mixing. It is known that
the \mt interchange or reflection symmetry when simultaneously imposed on
the charged leptons and the neutrinos do not lead to phenomenologically
viable leptonic mixing. In the case of \mt interchange symmetry, the obtained
forms of ${U_l}_L$ and $U_{\nu_L}$ given in Eq. (\ref{umutau}) imply that
the leptonic mixing matrix $U_{\rm PMNS}=U_{l_L}^\dagger U_{\nu_L}$ has
vanishing atmospheric and reactor mixing angles \cite{Joshipura:2005vy}.
This does not happen if one uses \mt reflection symmetry to forbid KM. But
in this case, one gets vanishing leptonic CP violation. This general result
can be shown following the arguments given in \cite{Grimus:2017itg} in a
slightly different context. In the case of \mt reflection symmetry,
$U_{l_L}$ and $U_{\nu_L}$ given in  Eq. (\ref{umutaur}) diagonalize 
$M_lM_l^\dagger$ and $M_\nu$ of Eq. (\ref{mutaur}), respectively. The $U_{
l_L}^{\rm HS}$ and $U_{\nu_L}^{\rm HS}$ satisfy
\be \label{slsnu}
S U_{l_L}^{\rm HS}=(U_{l_L}^{\rm HS})^*\,,~~S U_{\nu_L}^{\rm
HS}=(U_{\nu_L}^{\rm HS})^*\,.\ee
As a consequence,
\be \label{}
U_{\rm PMNS}^*= U_{l_L}^\dagger S^2 U_{\nu_L}=U_{l_L}^T  U_{\nu_L}^*=U_{\rm
PMNS}\,,\ee
and thus leads to a real $U_{\rm PMNS}$. Since $X_\nu$ in Eq. (\ref{umutaur}) is
trivial, the Dirac and Majorana phases vanish and there is no CP
violation in the lepton sector. $U_{\rm PMNS}$ matrix in this case is otherwise
general and allows arbitrary values of all the three mixing angles. If the
leptonic CP violation is found to be absent, then the \mt reflection symmetry
can provide an explanation of this and would also forbid KM parameters to all orders. However, one
would need to break this symmetry if non-trivial CP violation is to
be obtained.

Both the above discussed symmetries forbid KM parameters at the 1-loop level but fail in generating phenomenologically acceptable leptonic mixing. This can be
changed by generalizing the definition of \mt interchange or reflection
symmetries. It is assumed that these symmetries are symmetries of the Yukawa
interactions but get broken in  such a way that the leptonic mass matrices
$M_l$ and $M_\nu$ are invariant under different residual symmetries. The
idea of using different residual symmetries for the charged leptons and
neutrinos is extensively used in constraining leptonic mixing patterns
through discrete symmetries (see
\cite{Altarelli:2010gt,Smirnov:2011jv,King:2013eh,Ishimori:2010au} for
reviews).
Denoting these symmetries by $S_{l_{L,R}}$ and $S_{\nu_L}$, we demand
\be \label{genmtr}
S_{l_L}^\dagger M_lM_l^\dagger
S_{l_L}=(M_lM_l^\dagger)^*\,,~~S_{l_R}^\dagger M_l^\dagger M_l
S_{l_R}=(M_l^\dagger M_l)^*\,,~~S_{\nu_L}^T M_\nu S_{\nu_L}=M_\nu^*\,.\ee
The symmetry operators $S_{l_L}$, $S_{l_R}$ and $S_{\nu_L}$ are required to
constraint the diagonalizing matrices $U_{l_L}$, $U_{l_R}$ and $U_{\nu_L}$
such that each satisfy Eq. (\ref{mutaurgen}) needed to obtain vanishing KM parameters
at 1-loop. The most general solution of Eq. (\ref{mutaurgen}) can be written
as
\be \label{genu}
U=P\, U^{\rm HS}\, Q\,,\ee
where $P$ and $Q$ are diagonal phase matrices and $U^{\rm HS}$ is defined in
Eq. (\ref{hs}). Choosing the above form for $U_{l_L}$, $U_{l_R}$ and
$U_{\nu_L}$, we conveniently define
\be \label{ulunu}
U_{l_L}=P_{L} U^{\rm HS}_{L} Q_{L}\,,~~U_{l_R}=P_{R} U^{\rm HS}_{R}
Q_{R}\,,~~U_{\nu_L}=P_{\nu} U^{\rm HS}_{\nu} X_{\nu}\,, \ee
with 
\be \label{philphinu}
P_{L}={\rm Diag.}(1,e^{i \phi_{1L}},e^{i \phi_{2L}})\,,~~ P_{R}={\rm
Diag.}(1,e^{i \phi_{1R}},e^{i \phi_{2R}})\,,~~P_\nu= {\rm Diag.}(1,e^{i
\phi_{1\nu}},e^{i \phi_{2\nu}})\,, \ee
and $Q_{L}$, $Q_{R}$ are diagonal phase matrices. $X_\nu$ is a trivial
diagonal matrix with elements $\pm 1$ as before. The mass matrices which can
be diagonalized by the above unitary matrices have the form:
\be \label{diag_def}
M_\nu=\tilde{U}_{\nu_L}^* D_\nu \tilde{U}_{\nu_L}^\dagger\,,~~
M_lM_l^\dagger=\tilde{U}_{l_L} |D_l|^2 \tilde{U}_{l_L}^\dagger\,,~~
M_l^\dagger M_l=\tilde{U}_{l_R} |D_l|^2 \tilde{U}_{l_R}^\dagger\,.\ee
The above definitions together with Eqs. (\ref{ulunu}) imply
\be \label{sr}
S_{L}\equiv P_{L} S P_{L}\,,~~S_{R}\equiv P_{R} S P_{R}\,,~~S_{\nu }\equiv
P_\nu S P_\nu\ \ee
and satisfy Eq. (\ref{genmtr}). These symmetries thus represent the generalization of the \mt reflection symmetries. One recovers the \mt reflection symmetry if the phases satisfy
$\phi_{1a}=-\phi_{2a}$ for $a=L,R,\nu$. By construction, the generalized
symmetries lead to mixing matrices which assure vanishing KM parameters at  1-loop.
Moreover, the newly defined symmetries do not satisfy Eq. (\ref{slsnu}) used
in proving real $U_{\rm PMNS}$ as long as $P_L \neq P_\nu$ in Eq.
(\ref{ulunu}). One therefore gets a non-real and general $U_{\rm PMNS}$ which
allows Dirac CP violation. The Majorana phases still remain zero due to
triviality of $X_\nu$ in Eq. (\ref{ulunu}). The action of the above symmetries on the leptonic fields is given by
\beqa \label{mutaur_tr}
\mu_L^\prime & \rightarrow &  e^{i(\phi_{1L}+\phi_{2L})}\tau_L^{\prime
CP}\,,~\mu_R^\prime \rightarrow e^{i(\phi_{1R}+\phi_{2R})}\tau_R^{\prime
CP}\,,~\nu_{\mu L}^\prime \rightarrow 
e^{i(\phi_{1\nu}+\phi_{2\nu})}\nu_{\tau L}^{\prime CP}\,, \nonumber \\
\tau_L^\prime &\rightarrow &  e^{i(\phi_{1L}+\phi_{2L})}\mu_L^{\prime
CP}\,,~\tau_R \rightarrow e^{i(\phi_{1R}+\phi_{2R})} \mu_R^{\prime CP}\,,
~\nu_{\tau L}^\prime\rightarrow
e^{i(\phi_{1\nu}+\phi_{2\nu})}\nu_{\mu L}^{\prime CP}\,.\eeqa
The neutral current interactions given below in Eq. (\ref{L_NC}) are invariant
under these transformation if one also transforms the gauge fields as in Eq.
(\ref{mutaur2}). This forbids KM at the tree and 1-loop  level. But now the
charged current interactions do not remain invariant under these symmetries
when $P_L \neq P_\nu$ in Eq. (\ref{ulunu}). This would lead to kinetic
mixing at the two loop level in general.

One can analogously define generalisation of the \mt symmetry with similar
consequences. This is given by
\be \label{smutau}
\hat{S}_{L}\equiv P_LSP_L^*\,,~~\hat{S}_{R}\equiv
P_RSP_R^*\,,~~\hat{S}_{\nu}\equiv P_\nu S P_\nu^*~\ee
In this case Eq. (\ref{genmtr}) is replaced by
\be \label{genmt}
\hat{S}_{L}^\dagger  M_lM_l^\dagger
\hat{S}_L=M_lM_l^\dagger\,,~~\hat{S}_{R}^\dagger  M_l^\dagger M_l
\hat{S}_R=M_l^\dagger M_l\,,~~\hat{S}_{\nu}^T M_\nu \hat{S}_{\nu
}=M_\nu~.\ee 
Again, the neutral current couplings are invariant under this symmetry and
do not lead to KM at the tree and 1-loop level but the charged current
interactions violate this symmetry in general. 
The leptonic mixing matrix is quite general in this case and unlike in the
case of \mt interchange symmetry, one does not get the unwanted result of
vanishing $\theta_{23}$ and $\theta_{13}$.

\section{Kinetic mixing: General Considerations}
\label{sec:general}
As discussed in the previous section, the exact \mt interchange or reflection symmetry forbidding the KM in gauged \lmt is inconsistent with the observed lepton mixing pattern. One therefore needs to break these symmetries either in the charged lepton or in the neutrino sector. In the absence of such
symmetries, the KM gets generated at 1-loop level even if it is assumed to be absent at the tree level. In this section, we first derive a general formula for 1-loop induced kinetic and mass mixing in the SM extended with $U(1)_X$ gauge symmetry. We then discuss their implications for \lmt models.

Let $f^\prime_{aL}$ and $f^\prime_{aR}$ with $a=1,2,...,n$ be $n$ copies of
left and right-handed fermions with hypercharges $Y^\prime_{La}$ and
$Y^\prime_{Ra}$, respectively. The corresponding $U(1)_X$ charges are
$X^\prime_{La}$ and $X^\prime_{Ra}$. These $n$ copies include three generations of the SM fermions with $a=i,j=1,2,3$ and $(n-3)$ additional
fermions with $a=m=4,...,n$. The neutral current interactions between these
fermions and vector bosons of abelian symmetries are given by
\be\label{L_NC}
- {\cal L}_{\rm NC} = g_Y B_\mu \left(Y^\prime_{La} \overline{f}^\prime_{aL}
\gamma^\mu f^\prime_{aL}   + Y^\prime_{Ra} \overline{f}^\prime_{aR}
\gamma^\mu f^\prime_{aR} \right)+g^\prime Z^\prime_\mu \left(X^\prime_{La}
\overline{f}^\prime_{aL} \gamma^\mu f^\prime_{aL}+X^\prime_{Ra}
\overline{f}^\prime_{aR} \gamma^\mu f^\prime_{aR}\right)\,,
\ee
where $g_Y$ and $g'$ are the gauge couplings corresponding to $U(1)_Y$ and
$U(1)_X$ gauge groups\footnote{The hypercharge is normalized such that the
electric charge is $Q=T_3+Y'$ and $g_Y=\frac{e}{\cos\theta_W}$.},
respectively. All the $n$ fermions ${f^\prime_a}_{L,R}$ of a given charge 
and helicity mix among themselves. The mass basis, denoted by ${f_a}_L$ and
${f_a}_R$, is defined by
\be \label{basis_change}
{f'_a}_{L,R} = \left({\cal U}_{f_{L,R}}\right)_{ab}\, f_{b L,R}\,, \ee
where ${\cal U}_{f_{L,R}}$ are $n \times n$ matrices. Eq. (\ref{L_NC}) can
be rewritten in terms of the mass basis as
\be \label{L_NC_mass}
- {\cal L}_{\rm NC} = g_Y B_\mu \left({Y_L}_{ab}\overline{f}_{aL} \gamma^\mu
f_{bL}+{Y_R}_{ab} \overline{f}_{aR} \gamma^\mu f_{bR}\right) + g^\prime
Z^\prime_\mu \left({X_L}_{ab} \overline{f}_{aL} \gamma^\mu f_{bL}+{X_R}_{ab}
\overline{f}_{aR} \gamma^\mu f_{bR}\right)\,,\,
\ee
where the matrices $X_{L,R}$ and $Y_{L,R}$ denote gauge charges in the mass
basis. They are obtained as
\beqa\label{XY}
X_{L,R} &=& {\cal U}_{f_{L,R}}^\dagger\, X'_{L,R}\, {\cal U}_{f_{L,R}}\,,
\nonumber\\
Y_{L,R} &=& {\cal U}_{f_{L,R}}^\dagger\, Y'_{L,R}\, {\cal U}_{f_{L,R}}\,,
\eeqa
where $X'_{L,R}={\rm Diag.}({X'_{L,R}}_1, {X'_{L,R}}_2,...,{X'_{L,R}}_n)$
and $Y'_{L,R}={\rm Diag.}({Y'_{L,R}}_1, {Y'_{L,R}}_2,...,{Y'_{L,R}}_n)$.

The interactions  in Eq. (\ref{L_NC_mass}) can contribute to the mixing
between the $B$ and $Z'$ bosons at loop level through the vacuum
polarization effects. Denoting the amplitude of vacuum polarization as $i
\Pi^{\mu \nu}_{BZ'}(q^2)$, it is parametrized as
\be \label{PI_gen}
\Pi^{\mu \nu}_{BZ'}(q^2) = \left(g^{\mu \nu} q^2 - q^\mu q^\nu \right)\,
A_{BZ'}+ g^{\mu \nu}\, B_{BZ'}\,.
\ee
Here, the parameter $A_{BZ'}$ can be identified with KM while
$B_{BZ'}$ would give rise to mass mixing between the $B$ and $Z'$ bosons.
Within this framework, 1-loop computation of the vacuum polarization
diagrams  gives
\be \label{A_gen}
A_{BZ'} = \frac{g_Y g'}{4 \pi^2}  \left[-\frac{1}{6}{\rm Tr}(Y'_L X'_L +
Y'_R X'_R)\, E + \sum_{a,b} \left({Y_L}_{ab} {X_L}_{ba}+{Y_R}_{ab}
{X_R}_{ba}\right) b_2[m_a,m_b,q]\right]\,,
\ee
and
\beqa \label{B_gen}
B_{BZ'} &=& \frac{g_Yg'}{8 \pi^2} \left[\left( \frac{1}{2} {\rm
Tr}(D^2(\{Y_L,X_L\}+\{Y_R,X_R\})) - {\rm Tr}(Y_L D X_R D + Y_R D X_L
D)\right) E \right. \nonumber \\
&-&\sum_{a,b} \left({Y_L}_{ab} {X_L}_{ba} +{Y_R}_{ab} {X_R}_{ba}\right)
(m_a^2 b_1[m_a,m_b,q]+m_b^2 b_1[m_b,m_a,q]) \nonumber \\
&+& \left. \sum_{a,b} \left({Y_L}_{ab} {X_R}_{ba}+{Y_R}_{ab}
{X_L}_{ba}\right) m_a m_b\, b_0[m_a,m_b,q] \right] \,. \eeqa
Here, 
\be \label{E}
E=\frac{2}{\epsilon} -\gamma + \ln(4 \pi) - \ln(\mu^2)\,,
\ee
and $\epsilon= 4-d$ in the dimensional regularization scheme. The parameter $\mu$ is an arbitrary subtraction scale. The terms proportional to $E$ in both $A_{BZ'}$ and $B_{BZ'}$ are divergent in four dimensions. $m_a$ is the mass of $a^{\rm th}$ fermion and $D={\rm Diag.}(m_1,m_2,...,m_n)$. The loop integration functions $b_0$, $b_1$, $b_2$ are listed as Eq. (\ref{app_int}) in Appendix \ref{sec_app}.

$A_{BZ'}$ and $B_{BZ'}$ in Eqs. (\ref{A_gen},\ref{B_gen}) characterize
1-loop contributions to kinetic and mass mixing parameters. More explicitly,
at 1-loop:
\be \label{}
\sin \chi = (\sin \chi)_{\rm tree} + \sum_f A_{BZ'}\,,~~\delta M^2 = (\delta
M^2)_{\rm tree} + \sum_f B_{BZ'}\,,
\ee
where the sum is over different kind of fermions present in the underlying
model. In the presence of non vanishing $\sin\chi$ or $\delta M^2$, the
gauge bosons $B$ and $Z'$ mix and their mass eigenstates $\tilde{B}$ and
$\tilde{Z}'$ can be obtained as \cite{Babu:1996vt}:
\beqa \label{gauge_massbasis}
\tilde{B} &=& \cos\xi\, B + \sin(\xi + \chi)\, Z'\,, \nonumber \\
\tilde{Z}' &=& -\sin\xi\, B +\cos(\xi + \chi)\, Z' \,,\eeqa
where
\be \label{xi}
\tan 2\xi = \frac{-2 \cos \chi\, (\delta M^2 - M_B^2 \sin
\chi)}{M_{Z^\prime}^2 - M_B^2 \cos 2\chi + 2\,  \delta M^2 \sin\chi}\ee
The angle $\xi$ is a phenomenologically useful parameter which quantifies the
overall effect of gauge boson mixing. Non-zero value of $\xi$ gives rise to
deviation in the neutral current couplings associated with the $Z$ boson
from their values predicted in the SM\footnote{For mixing with the standard
$Z$ boson instead of $B$, the mixing angle $\xi$ is obtained by the
replacements $\sin\chi \to - \sin\theta_W \sin\chi$ and $M_B \to M_Z$ in Eq.
(\ref{xi}).}.

 As seen from Eq. (\ref{A_gen}), the divergent part in $A_{BZ'}$ vanishes if the fermions ${f_a}_{L,R}$ have universal hypercharges and ${\rm Tr}(X'_{L,R}) = 0$. Consequently,  finiteness of $A_{BZ'}$ at 1-loop follows from the charge assignments of fermions and it does not require additional symmetry\footnote{Additional symmetries can play role in finiteness of KM at higher loops \cite{delAguila:1995rb}.}. This is the case in
the standard \lmt models used in many works for generating KM at 1-loop through the charged lepton
exchanges. One is left with finite and nonzero 1-loop contribution to $\sin\chi$ in this case for non-vanishing diagonal elements in $X_L$ and $X_R$. For example, if the charged lepton mass matrix $M_l$ is diagonal then one obtains the well-known \cite{Holdom:1985ag} result 
\be \label{kmch}
A_{BZ'} \approx  -\frac{ g_Yg'}{16 \pi^2}\, \ln \frac{m_\mu^2}{m_\tau^2}\,
\ee
from Eq. (\ref{A_gen}) in the limit $q^2 \ll m_\mu^2$. If \mt interchange or
reflection symmetry is imposed on $M_l$ then the resulting condition Eq.
(\ref{mutaurgen}) leads to vanishing diagonal elements in $X_{L,R}$ which
gives $A_{BZ'}=0$. Finiteness of $A_{BZ'}$ obtained at 1-loop may not hold
at higher loops if there is no \mt interchange like symmetry or its breaking is
hard.

The divergent part in $B_{BZ'}$ does not vanish in general. If fermions
$f_{a L,R}^\prime$  carry  universal hypercharges $Y'_{L,R}$, then their
contribution $B_{BZ^\prime}$ can be written as
\be \label{divB}
\frac{g_Y g^\prime}{8 \pi^2}\,(Y'_L-Y'_R)\, {\rm Tr} (D^2 (X_L - X_R))\,
E\,.\ee
This piece vanishes only under the specific circumstances: (a) universal
masses $m_a$ since ${\rm Tr}(X_{L,R})=0$, (b) vectorial hypercharge, i.e.
$Y'_L=Y'_R$, (c) vectorial $Z'$ current, i.e. $X_L=X_R$ or (d) generalised
\mt symmetry as defined by Eq. (\ref{mutaurgen}) for which diagonal elements
of $X_{L,R}$ vanish individually. None of these conditions are automatically
satisfied for the charged leptons with a general non-Hermitian mass matrix
$M_l$. Only if $M_l$ is Hermitian or possesses one of the symmetries
discussed in the previous sections, the divergent piece in $B_{BZ'}$
vanishes. Otherwise, the 1-loop contribution to $B_{BZ'}$ is divergent. Thus,
in spite of finite and calculable contribution from Eq. (\ref{A_gen}), the
charged lepton contribution to the $Z$-$Z'$ mixing  remains incalculable.
Similarly, for the neutrino sector, if the fields ${f^\prime_L}_i$, ${f^\prime_R}_i$ represent the standard light Majorana neutrinos with $f^\prime_R=C \overline{f}^{\prime T}_L $ then $Y'_L = - Y'_R$ and $(X_L)_{ii}=-(X_R)_{ii}$ in Eq. (\ref{divB}). As a consequence, the light Majorana
neutrino contribution to $B_{BZ'}$ is always divergent unless they are
degenerate or the neutrino mass matrix is invariant under one of the
symmetries discussed earlier. Since parameter $B_{BZ'}$ contributes to
$\xi$, the resultant $Z$-$Z'$ mixing remains divergent and incalculable in
the minimal set up with general mass matrices for the charged leptons and
neutrinos. 

There exists one specific scenario for which the 1-loop expression of KM parameter
as given in Eq. (\ref{kmch}) holds, $Z$-$Z'$ mixing is calculable, and the
neutrino mixing is also consistent with the current results. This
corresponds to assuming unbroken \lmt symmetry for the charged leptons and
\mt reflection symmetry for the neutrino sector. In this case, neutrinos do
not contribute to $A_{BZ'}$ and $B_{BZ'}$ at 1-loop as discussed in the
previous section, the charged lepton contribution to $B_{BZ'}$ vanishes and
their contribution to $A_{BZ'}$ is finite and given by Eq. (\ref{kmch}).
Unbroken \mt reflection symmetry in the neutrino sector predicts maximal
atmospheric mixing angle as well as maximal Dirac CP violation.

The divergent part of $B_{BZ'}$ can be renormalized by introducing suitable counter term as the \mt symmetry is already broken in the effective framework. Hence, there is no reason for such counter terms to be not present in the theory. However, in the full ultraviolet completion of the model in which the \mt interchange or reflection symmetry is restored, the divergences in the kinetic and mass mixing terms must not arise. In these models, the KM parameters are calculable, and its origin can be linked with the mechanism of \mt symmetry breaking. We provide two characteristically different frameworks as the concrete realization of this statement in the next section.

\section{Models of calculable kinetic mixing}
\label{sec:model}

We consider (A) the standard seesaw model and (B) a model with vectorlike
charged leptons in which the underlying \mt symmetry is broken spontaneously
or softly leading to finite $\sin\chi$ and $\delta M^2$ at 1-loop. Both of
these represent special cases of the general formalism discussed in the last
section.

\subsection{Kinetic mixing in the standard seesaw model}
\label{sec:model_I}
The model is the standard seesaw model augmented with a gauge \lmt symmetry
and a \mt symmetry. Breaking of \lmt occurs spontaneously through $SU(2)_L
\times U(1)_Y$ singlet fields. As a consequence, parameter $\delta M_Z^2$
does not get generated at tree level even after breaking of the \lmt and SM
gauge symmetries. The charged lepton masses in the model are characterized
by the following mass Lagrangian
\be \label{Lf}
{-\cal L}_{m}^l = \left( \ba{ccc} \overline{e}'_L & \overline{\mu}'_L &
\overline{\tau}'_L \ea \right)\, M_l\, \left( \ba{c} e'_R \\ \mu'_R \\
\tau'_R \ea \right)+{\rm h.c}~, \ee 
with
\be \label{ml}
M_l= v \left( \ba{ccc} \lambda^l_{ee} & \lambda^l_{e \mu} & \lambda^l_{e
\mu} \\
\lambda^l_{\mu e} & \lambda^l_{\mu \mu} & \lambda^l_{\mu \tau} \\
\lambda^l_{\mu e} & \lambda^l_{\mu \tau} & \lambda^l_{\mu \mu} \ea
\right)\,.
\ee
Here, $v$ is the vacuum expectation value (VEV) of the standard model doublet
assumed neutral under \lmt and \mt symmetry. The off-diagonal couplings
$\lambda^l_{\mu \tau}$, $\lambda^l_{e \mu}$, $\lambda^l_{\mu e}$ can be
regarded as VEVs of the spurion fields with \lmt charge $2$, $-1$ and $1$,
respectively. It is assumed that these fields break \lmt symmetry
spontaneously but do not break the \mt interchange symmetry which leads to
the above form of $M_l$. Similarly, the Dirac neutrino mass matrix is also assumed to be invariant under the \mt interchange symmetry and has the form
\be \label{mD}
m_D=v \left( \ba{ccc} \lambda^D_{11} & \lambda^D_{12} & \lambda^D_{12} \\
\lambda^D_{21} & \lambda^D_{22} & \lambda^D_{23} \\
\lambda^D_{21} & \lambda^D_{23} & \lambda^D_{22} \ea
\right)\,.\ee
Non-zero off-diagonal couplings in $m_D$ arise because of the spontaneous breaking of \lmt symmetry.

Unlike in the case of $M_l$ and $m_D$, the \mt symmetry is assumed to be broken
by the Majorana masses of RH neutrinos. This can be achieved
by introducing an appropriately charged spurions field whose VEV break both the
\mt and \lmt symmetries spontaneously. This allows a completely general form for the RH
neutrino mass matrix $M_R$ and thereby leads to a general lepton mixing
matrix. We shall work out the radiatively generated KM parameters for this
general matrix. Special cases can be obtained by restricting the structure
of $M_l$ and $M_R$. Specific neutrino mass structures and associated
phenomenology has been discussed in a number of papers
\cite{Ma:2001md,Choubey:2004hn,Dev:2017fdz,Asai:2018ocx} in the context of the \lmt symmetry.
The neutrino mass Lagrangian is defined as
\be \label{Lf_nu}
{-\cal L}_{m}^\nu = \frac{1}{2}\, n^{\prime T}_L\, C {\cal M}_\nu\, 
n^{\prime}_L\,+{\rm h.c}~, \ee 
where $n_L^\prime\equiv(\nu_L^\prime,(\nu_{R}^\prime)^c)^T$ is a six
dimensional column vector of the left-handed fields. The right-handed
components are analogously defined as $n_R^\prime = (n_L^\prime)^c =
((\nu_L^\prime)^c,\nu_{R}^{\prime})^T$. The $6\times 6$ Majorana neutrino
mass matrix is
\be \label{seesaw}
{\cal M}_\nu=\left(\ba{cc}
0 & m_D^T\\
m_D & M_R\ea \right)\,.\ee
Six neutrino mass eigenstates are then obtain using the following unitary
transformations:
\be \label{masses}
n_L^\prime={\cal U}\, n_L\,,~~n_R^\prime={\cal U}^*\, n_R\,.\ee
The chiral components $n_{L,R}$ of six Majorana  mass eigenstates  can be
identified with the light and heavy neutrino mass eigenstates as: $n_{iL}=
\nu_{iL}$, $n_{(i+3)L}= (\nu_{iR})^c$ and $n_{iR}= (n_{i L})^c$. The mixing
matrix ${\cal U}$ is required  to satisfy
\be \label{seesawdia}
{\cal U}^T\, {\cal M}_\nu\, {\cal U}= {\cal D}_\nu \equiv {\rm
Diag.}(m_{\nu_i},M_i)~\ee
where $m_{\nu_i},M_i$ are respectively light and heavy neutrino masses.

The neutral current interactions of neutrinos in the $n^\prime_L$  basis are
given by
\be \label{chiBprime1}
-{\cal L}_{\rm NC} = g_YB_\mu\, \tilde{Y}_a\overline{n}_{aL}^\prime
\gamma^\mu n_{aL}^\prime+g^\prime
Z^\prime_\mu\tilde{X}_a\overline{n}_{aL}^\prime \gamma^\mu n_{aL}^\prime
\,,\ee
where $\tilde{Y} = -\frac{1}{2}{\rm Diag.}(1,1,1,0,0,0)$ and $\tilde{X} =
{\rm Diag.}(0,1,-1,0,-1,1)$.
Using the Majorana property $\overline{n}_{aL}^\prime \gamma^\mu
n_{aL}^\prime=-\overline{n}_{aL}^{c\prime} \gamma^\mu
n_{aL}^{c\prime}=-\overline{n}_{aR}^\prime \gamma^\mu n_{aR}^\prime$, the above equation can be cast in the following form:
\be \label{chiBprime}
-{\cal L}_{\rm NC}=\frac{g_Y}{2}B_\mu\,
\tilde{Y}_a\left(\overline{n}_{aL}^\prime \gamma^\mu n_{aL}^\prime -
\overline{n}_{aR}^\prime \gamma^\mu n_{aR}^\prime\right) +
\frac{g^\prime}{2} Z^\prime_\mu\, \tilde{X}_a\left(\overline{n}_{aL}^\prime
\gamma^\mu n_{aL}^\prime - \overline{n}_{aR}^\prime \gamma^\mu
n_{aR}^\prime\right)\,.\ee
Following the arguments presented between Eqs. (\ref{L_NC}) and
(\ref{XY}) for the general case, we obtain in the mass basis
\be \label{chiBZ}
-{\cal L}_{\rm NC}=\frac{g_Y}{2}B_\mu\, \overline{n}\gamma^\mu\left({\cal
U}^\dagger \tilde{Y} {\cal U} P_L-{\cal U}^T \tilde{Y} {\cal U}^*
P_R\right)n+
\frac{g^\prime}{2}Z_\mu^\prime\,  \overline{n}\gamma^\mu\left({\cal
U}^\dagger \tilde{X} {\cal U}P_L-{\cal U}^T \tilde{X} {\cal U}
P_R\right)n\,.\ee
Eq. (\ref{chiBprime}) can be seen as special case of the general expression
Eq. (\ref{L_NC_mass}) with the identification
\beqa \label{xyseesaw}
(Y_L)_{ab}= \frac{1}{2} \left({\cal U}^\dagger \tilde{Y} {\cal
U}\right)_{ab}&\,,~~&(X_L)_{ab}= \frac{1}{2}\left({\cal U}^\dagger \tilde{X}
{\cal U}\right)_{ab}\,,\nonumber \\
(Y_R)_{ab}=-\frac{1}{2} \left({\cal U}^T \tilde{Y} {\cal
U}^*\right)_{ab}&\,,~~&(X_R)_{ab}=-\frac{1}{2}\left({\cal U}^T \tilde{X}
{\cal U}^*\right)_{ab}\,.\eeqa
The above expressions also follow directly from Eq. (\ref{XY}) by noting
that (i) the left and right handed mixing matrices are related as ${\cal
U}_L={\cal U}_R^*\equiv {\cal U}$ (see Eq. (\ref{masses})), and (ii) the 
$U(1)$ charges of $n_R$ and $n_L$ are opposite to each other. One can
use Eq. (\ref{xyseesaw}) to directly obtain KM parameters in the present
case. In this we closely follow the treatment of radiative corrections given
in \cite{Pilaftsis:1991ug,Korner:1992an,Grimus:2002nk}.

It is trivial to see from the comparison with Eq. (\ref{A_gen}) that the
divergent part in $A_{BZ'}$ vanishes for the present case as ${\rm
Tr}(\tilde{X} \tilde{Y})=0$. To show finiteness of $B_{BZ'}$, it is useful
to decompose  ${\cal U}$ as
\be \label{calu}
{\cal U}=\left ( \ba{c}
V_{L}\\
V_R^*\\ \ea \right)~\ee
in terms of $3\times 6$ matrices  $V_L$ and $V_R$. The matrices $X_{L,R}$
and $Y_{L,R}$ are then given by
\be \label{xlxr}
X_L=-X_R^*=\frac{1}{2}(V_L^\dagger X_3 V_L-V_R^T X_3
V_R^*)\,,~~Y_L=-Y_R^*=-\frac{1}{4}V_L^\dagger V_L\,.\ee
Here, $X_3={\rm Diag.}(0,1,-1)$. Eq. (\ref{seesawdia}) and unitarity of ${\cal
U}$ can be used to derive the relations
\be \label{conditions1}
V_LV_L^\dagger={\bf 1}_{3\times 3}\,,~~V_LV_R^T={\bf 0}_{3\times
3}\,,~~V_RV_R^\dagger={\bf 1}_{3\times 3}\,, \ee
\be \label{conditions2}
V_L{\cal D}_\nu V_L^T={\bf 0}_{3\times 3}\,,~~V_R {\cal D}_\nu
V_L^\dagger=m_D\,,~~ V_R{\cal D}_\nu V_R^T=M_R\,.\ee
Here, ${\bf 1}_{3\times 3}$ and ${\bf 0}_{3\times 3}$ respectively denote
the $3\times 3$ identity and null matrix. The expressions for KM
parameters follow by substituting Eq. (\ref{xyseesaw}) in the general
formula Eqs. (\ref{A_gen},\ref{B_gen}). Finiteness of $B_{BZ'}$ then follows
from the following identities
\be \label{cseesaw1}
\frac{1}{2} {\rm Tr}\left({\cal D}_\nu^2(\{Y_L,X_L\}+\{Y_R,X_R\})\right) = -\frac{1}{4} {\rm Tr} \left(m_D^\dagger m_D X_3\right)\,,\ee
\be\label{cseesaw2}
{\rm Tr} \left(Y_L {\cal D}_\nu X_R {\cal D}_\nu + Y_R {\cal D}_\nu X_L
{\cal D}_\nu \right)=-\frac{1}{4} {\rm Tr} \left(m_D m_D^\dagger X_3 \right)\,.\ee
We have used the definition of $X_L$, $Y_L$ and Eqs. (\ref{conditions1},\ref{conditions2}) in proving above equations. The divergent part in $B_{BZ'}$ vanishes for \mt symmetric $m_D$ given in Eq. (\ref{mD}) as both the Eqs. (\ref{cseesaw1},\ref{cseesaw2}) vanish individually. The finite parts can be written as
\be \label{A_seesaw}
A_{BZ'} = \frac{g_Y g'}{2 \pi^2} \sum_{a,b} {\rm Re}\left[{Y_L}_{ab}
{X_L}_{ba}\right] b_2[m_a,m_b,q]\,,\ee
\beqa \label{B_seesaw}
B_{BZ'} &=& -\frac{g_Yg'}{4 \pi^2} \sum_{a,b} \Big( {\rm Re}\left[{Y_L}_{ab}
{X_L}_{ba}\right]  (m_a^2 b_1[m_a,m_b,q]+m_b^2 b_1[m_b,m_a,q]) \Big.
~,\nonumber\\
&+& \Big.{\rm Re}\left[{Y_L}_{ab} {X_L^*}_{ba}\right] m_a m_b\,
b_0[m_a,m_b,q] \Big)\,. \eeqa

The above considerations are valid in general seesaw model without taking the standard limit $m_D \ll M_R$. We now consider this limit in order to further simplify the finite contributions to KM parameters. ${\cal U}$ can be written as
\be \label{useesaw} {\cal U}=\left( \ba{cc} 
1-\frac{1}{2}\rho\rho^\dagger&-\rho \\
\rho^\dagger &1-\frac{1}{2}\rho^\dagger\rho \\
\ea\right) \left(\ba{cc}
K_L&0\\
0&K_R\\ \ea \right)~.\ee
In the seesaw limit,  $\rho^\dagger\approx -M_R^{-1} m_D$, while $K_L$ and $K_R$ are $3\times 3$ matrices which diagonalize the light and heavy neutrino matrices $m_\nu=-m_D^TM_R^{-1}m_D$ and $M_R$ respectively. Parameters $A_{BZ'}$ and $B_{BZ'}$ can be simplified in a special case of the second and third generations.  Further simplification can be achieved if $m_D$ is assumed to be invariant under \lmt and thus diagonal. In this case, it is  proportional to $2\times 2$ identity matrix and explicitly
$$m_D\equiv m\, {\bf 1}_{2\times 2}.$$
The light neutrino mass matrix is then given by $-m^2 M_R^{-1}$ and therefore the matrices $K_L$ and $K_R$ are related as $K_R=K_L^*$. In this case, $V_L$ and $V_R$ defined in Eq. (\ref{calu}) simplify to
\be \label{vlvrseesaw}
V_L=(K_L,-\rho K_R)\approx( K_L,mK_L D_R^{-1})\,,~~ V_R^*=(\rho^\dagger
K_L,K_R) \approx (-m K_L^* D_R^{-1},K_L^*)\,,\ee
where $D_R={\rm Diag.}(M_2,M_3)$. We parametrise $2\times 2$ matrix $K_L$
as
\be \label{KL}
K_L = \left( \ba{cc} \cos\theta & \sin\theta \\  -\sin\theta & \cos\theta
\ea\right)\,.\ee
This together with Eq. (\ref{vlvrseesaw})  determine the parameters 
$(X_L)_{ab}$, $(Y_L)_{ab}$ and lead to
\beqa \label{appAB}
A_{BZ'}  &\approx & \frac{g_Y g^\prime}{{16 \pi^2}} \cos 2\theta
\Big(b_2(m_{\nu_3},m_{\nu_3},q) - b_2(m_{\nu_2},m_{\nu_2},q) \Big. \nonumber
\\
& + & \frac{m^2}{M_3^{2}} \left(4 b_2(M_3,m_{\nu_3},q) - b_2(M_3,M_3,q)  - 3 b_2(m_{\nu_3},m_{\nu_3},q) \right)
\nonumber \\
& - & \left. \frac{m^2}{M_2^{2}} \left(4 b_2(M_2,m_{\nu_2},q) - b_2(M_2,M_2,q)   - 3 b_2(m_{\nu_2},m_{\nu_2},q)\right)
+{\cal O}\left( \frac{m^4}{M_{2,3}^4}\right)\right)\,, \nonumber \\
B_{BZ'} &\approx & \frac{g_Y g^\prime}{{16 \pi^2}} \cos 2\theta\, m^2
\Big(b_0(M_3,M_3,q) - 2 b_1(M_3,m_{\nu_3},q) \Big. \nonumber \\
& - & \left. b_0(M_2,M_2,q) + 2b_1(M_2,m_{\nu_2},q) + {\cal O}\left(\frac{m^2}{M_{2,3}^2} \right) \right)\,,\eeqa

In the limit $m_{\nu_{2,3}}^2 \ll |q^2| \ll M_i^2$, using the approximate solutions of integration functions provided in Eqs. (\ref{app_int_1},\ref{app_int_2},\ref{app_int_3}) in Appendix, we obtain 
\beqa \label{numab}
A_{BZ'} & \approx & \frac{g_Y g^\prime}{16 \pi^2} \cos2 \theta \left(
\frac{\Delta_{\rm atm}}{-q^2}- \left(\frac{m^2}{M_2^2}-\frac{m^2}{M_3^2}\right)\left(\frac{5}{18}-\frac{1}{2} \ln\frac{-q^2}{\mu^2}\right) \right. \nonumber \\
& - & \left. \frac{m^2}{2 M_2^2} \ln \frac{M_2^2}{\mu^2} +
\frac{m^2}{2 M_3^2} \ln \frac{M_3^2}{\mu^2}\right)~,\nonumber\\
B_{BZ'} & \approx & - \frac{g_Y g^\prime}{32\pi^2} \cos 2\theta\, q^2
\left(\frac{m^2}{M_2^2}-\frac{m^2}{M_3^2}\right)\,,\eeqa
 at the leading order in $m^2/M^2$. The first term in $A_{BZ'}$ corresponds to the contribution from the effective
light neutrino mass matrix and the other two contributions of ${\cal
O}(\frac{m^2}{M^2})$ arise due to light heavy  neutrino mixing. All these
contributions vanish in the limit of the RH neutrino masses going to
infinity. 

The parameter $\theta$ in Eq. (\ref{numab}) is the neutrino part of the
atmospheric mixing angle. The charged lepton contribution to it is maximal
because of \mt symmetry of  $M_l$ and one thus gets
$\theta_{23}=\theta-\frac{\pi}{4}$. One therefore requires small $\theta$
and hence almost diagonal $M_R$. The RH neutrino masses $M_{2,3}$ in this
case are directly linked to the \lmt breaking scale. As a consequence, the
$Z^\prime$ mass would be similar to the RH neutrino masses unless $g^\prime$
is
very small. Light $Z^\prime$ is still a possibility if the RH neutrino mass
scale is around TeV, {\it e.g.}, $M_2,M_3\sim {\rm TeV}$ and $g^\prime\sim
10^{-3}$ would give $M_{Z^\prime}\sim {\rm GeV}$. The KM is still
suppressed by the light neutrino masses. For $M_{2,3}\sim {\rm TeV}$ and $-q^2\sim
{\rm MeV}^2$, the
dominant contribution to $A_{BZ'}$ comes from the last two terms in Eq.
(\ref{numab}) and is given by 
$$A_{BZ'} \sim  g_Y g^\prime ~ 3\cdot 10^{-16} \left(\frac{{\rm TeV}}{M_3}
\right)~.$$

There can be additional contributions to KM from the Higgs
sector. Such contributions would vanish in the exact \mt symmetric limit.
The \mt breaking in our case comes from the right-handed neutrino masses, 
which could be explicit or induced through singlet VEVs. But singlet fields do not directly couple to $Z$ and cannot induce $Z-Z'$ mixing. There can be indirect coupling through the quartic interaction $\lambda \eta^\dagger \eta\phi^\dagger\phi$ of the singlet field $\eta$ with the
$SU(2)$ doublet $\phi.$ This induced coupling of $\eta$ to $Z$ will be
suppressed by $\frac{\vev{\phi}}{\vev{\eta}}$ and the resulting mixing would
 also be suppressed.
 
 In the above example, we have assumed \mt symmetric $m_D$
which leads to finite $B_{BZ'}$ at the 1-loop. Alternatively, the
1-loop divergences in $B_{BZ'}$ also vanish if $m_D$ is not \mt symmetric
but it possesses unbroken \lmt symmetry. In this case, the finiteness of 
$B_{BZ'}$ follows from cancellation between the contributions
(\ref{cseesaw1}) and (\ref{cseesaw2}) in Eq. (\ref{B_gen}) because of
diagonal $m_D$. This case is phenomenologially more important since it
leads to a non-vanishing contribution even when the RH neutrino
masses are taken to infinity. We discuss an explicit example which
shows this. Assume again two generations with 
\beqa \label{mdmr}
m_D=\left(\ba{cc}
m_2&0\\
0&m_3\\ \ea\right) &,&M_R=\left(\ba{cc}
M_2&0\\
0&M_3\\ \ea\right)~.\eeqa
$K_{L,R}$ defined in Eq. (\ref{useesaw}) are $2\times 2$ identity matrices
and $\rho=-m_D M_R^{-1} $ is also diagonal. The $2\times 4$ matrices
$V_L$, $V_R$ can be obtained in this case from  Eq. (\ref{vlvrseesaw}) and one
can work out the resulting $4\times 4$ matrices $Y_{L,R}$, $X_{L,R}$ using
Eq. (\ref{xlxr}). This leads through Eqs. (\ref{A_seesaw},\ref{B_seesaw}) to
the following expressions for
$A_{BZ'}$ and $B_{BZ'}$:
\beqa
A_{BZ'} &\approx & \frac{g_Y g^\prime}{16\pi^2} \Big( b_2(m_{\nu_3},m_{\nu_3},q)-b_2(m_{\nu_2},m_{\nu_2},q) \Big.~ \nonumber\\
& +&\frac{m_3^2}{M_3^2}\left(4 b_2(M_3,m_{\nu_3},q)-b_2(M_3,M_3,q)-3 b_2(m_{\nu_3},m_{\nu_3},q) \right) \nonumber \\
&- & \left. \frac{m_2^2}{M_2^2}\left(4 b_2(M_2,m_{\nu_2},q)-b_2(M_2,M_2,q)-3 b_2(m_{\nu_2},m_{\nu_2},q) \right) +{\cal O} \left( \frac{m_{2,3}^4}{M_{2,3}^4}\right) \right)\, \nonumber \\
B_{BZ'} &\approx & \frac{g_Yg^\prime}{16\pi^2}\Big(m_3^2(b_0(M_3,M_3,q)-2 b_1(M_3,m_{\nu_3},q)) \Big. \nonumber \\ 
& - & \left. m_2^2(b_0(M_2,M_2,q)-2 b_1(M_2,m_{\nu_2},q))  + {\cal O}\left(\frac{m_{2,3}^2}{M_{2,3}^2} \right)\right)~,\eeqa
In the limit $m_{\nu_{2,3}}^2\ll |q^2|\ll M_{2,3}^2$, we obtain
\beqa
A_{BZ'}& \approx & \frac{g_Yg^\prime}{16\pi^2} \left( \frac{\Delta_{\rm atm}}{-q^2} - \left(\frac{m_2^2}{M_2^2}-\frac{m_3^2}{M_3^2} \right) \left(\frac{5}{18} -\frac{1}{2} \ln\frac{-q^2}{\mu^2}\right)-\frac{m_2^2}{2M_2^2} \ln\frac{M_2^2}{\mu^2}+\frac{m_3^2}{2M_3^2} \ln\frac{M_3^2}{\mu^2})\right)~,\nonumber\\
B_{BZ'}&\approx & -\frac{g_Yg^\prime}{32\pi^2}\left(3(m_2^2-m_3^2) + q^2 \left(\frac{m_2^2}{M_2^2}-\frac{m_3^2}{M_3^2}\right)\right)\,.\eeqa
Unlike in the previous case, there is a finite non-decoupling
contribution in $B_{BZ'}$ which does not vanish when the right handed
neutrino masses are taken to infinity. This contribution is proportional to the
amount of $\mu$-$\tau$ breaking, $m_2^2-m_3^2$, in $m_D$. $A_{BZ'}$ is still
seen to vanish when the RH neutrino masses go to infinity.

\subsection{Kinetic mixing in a model with vectorlike charged leptons}
\label{sec:model_II}
In this case, the effective $3 \times 3$ Majorana\footnote{Although we
assume neutrinos as Majorana fermions, the same results are obtained if they
are Dirac fermions.} neutrino mass matrix $M_\nu$ is assumed to be invariant
under \mt interchange symmetry. It is explicitly given as
\be \label{}
M_\nu = v_\nu \left( \ba{ccc} \lambda^\nu_{ee} & \lambda^\nu_{e\mu} &
\lambda^\nu_{e\mu}\\\lambda^\nu_{e\mu} & \lambda^\nu_{\mu \mu} &
\lambda^\nu_{\mu \tau}\\\lambda^\nu_{e\mu} & \lambda^\nu_{\mu \tau} &
\lambda^\nu_{\mu \mu} \ea  \right)\,. \ee
The couplings $\lambda^\nu_{e\mu}$, $\lambda^\nu_{\mu \mu}$ can be seen as
spurions which break \lmt symmetry spontaneously but preserve the \mt
interchange symmetry. Because of the later, the neutrinos by themselves do
not induce the KM between the $B$ and $Z'$ bosons.

The charged lepton sector is extended by a pair of vectorlike leptons,
$f'_4$ and $f'_5$, singlet under $SU(2)_{\rm L}$ and with hypercharge $-1$.
Under the gauged \lmt symmetry, $f'_4$ and $f'_5$ have charges $+1$ and
$-1$, respectively. Further, $f'_4$ and $f'_5$ get interchanged under the
\mt symmetry in addition to the transformations defined in Eq.
(\ref{mutau}). After the spontaneous breaking of \lmt and electroweak
symmetry, the charged lepton mass term in the Lagrangian is given by
\be  \label{Lf_cl}
{-\cal L}_{m}^l = \left( \ba{ccccc} \overline{e}'_L & \overline{\mu}'_L &
\overline{\tau}'_L & {\overline{f}'_4}_L & {\overline{f}'_5}_L\ea \right)\,
{\cal M}_l\, \left( \ba{c} e'_R \\ \mu'_R \\ \tau'_R \\ {f'_4}_R  \\
{f'_5}_R \ea \right)+{\rm h.c}~, \ee where
\be 
{\cal M}_l= \left( \ba{cc} (M_l)_{3 \times 3} & (m_l)_{3\times 2} \\ 
(\tilde{m}_l)_{2\times 3} & (M_f)_{2 \times 2} \ea \right)\,.  \ee
The matrix $M_l$ is invariant under \mt symmetry and  has the same form as in Eq.
(\ref{ml}). The explicit forms of the matrices $m$ and $\tilde{m}$ are
\be 
m_l= \left(\ba{cc} m_{e4} & m_{e4} \\ m_{\mu 4} & m_{\mu 5} \\ m_{\mu 5} &
m_{\mu 4} \ea \right)\,,~~~ \tilde{m}_l= \left(\ba{ccc} m_{4e} & m_{4 \mu} &
m_{4 \tau} \\ m_{4 e} & m_{4 \tau} & m_{4 \mu} \ea \right)\,,\ee
Both $m$ and $\tilde{m}$ are invariant under \mt interchange symmetry. The
mass terms $m_{e4}$, $m_{4e}$, $m_{\mu 5}$ and $m_{4 \tau}$ are spurious
which break the \lmt symmetry spontaneously. We assume general form for
matrix $M_f$ which breaks the \mt interchange symmetry softly unless
$(M_f)_{11} = (M_f)_{22}$ and $(M_f)_{12} =(M_f)_{21}$. This soft breaking
of \mt symmetry in $M_f$ leads to breaking of the same symmetry in the
effective theory obtained after integrating out the vectorlike charged
leptons. Therefore, the KM between $B$ and $Z'$ gets generated
at 1-loop in this setup.

The five mass eigenstates of the charged leptons are obtained using the
following unitary transformation.
\be \label{}
\left( \ba{c} e'_{L,R} \\ \mu'_{L,R} \\ \tau'_{L,R} \\ {f'_4}_{L,R}  \\
{f'_5}_{L,R} \ea \right) = {\cal U}_{L,R} \, \left( \ba{c} e_{L,R} \\
\mu_{L,R} \\ \tau_{L,R} \\ {f_4}_{L,R}  \\ {f_5}_{L,R} \ea \right)\, \ee 
such that 
\be \label{ml-diag}
{\cal U}_L^\dagger\, {\cal M}_l\, {\cal U}_R = {\rm Diag.}\left( m_e, m_\mu,
m_\tau, m_4, m_5\right) \equiv {\cal D}_l\,. \ee
For simplification, the $5 \times 5$ unitary matrices ${\cal U}_{L,R}$ can
be represented as
\be \label{UVdef}
{\cal U}_{L,R} = \left( \ba{c} U_{L,R} \\ V_{L,R} \ea \right)\,, \ee
where $U_{L,R}$  and $V_{L,R}$ are matrices of dimensions $3 \times 5$ and
$2 \times 5$ respectively. The unitarity of ${\cal U}_{L,R}$ and the
relation in Eq. (\ref{ml-diag}) can be used to obtain the following
relations:
\be \label{rel1_cl}
U_{L,R} U_{L,R}^\dagger = {\bf 1}_{3\times 3}\,,\,\, V_{L,R} V_{L,R}^\dagger
= {\bf 1}_{2\times 2}\,,\,\, U_{L,R} V_{L,R}^\dagger = {\bf 0}_{3\times
2}\,, \ee
\be \label{rel2_cl}
U_L {\cal D}_l U_R^\dagger = M_l\,,\,\, U_L {\cal D}_l V_R^\dagger = m_l\,,
\,\, V_L {\cal D}_l U_R^\dagger = \tilde{m}_l\,,\,\, V_L {\cal D}_l
V_R^\dagger = M_f\,. \ee

We now discuss the KM between $B$ and $Z'$ bosons induced at one
loop within this setup. The general formalism developed in section III can be
straight forwardly used to compute such mixing. The fermionic currents
associated with $Z'$ and $B$ bosons in this framework can be read from Eqs.
(\ref{L_NC}) with $f_a = \{e,\mu,\tau, f_4, f_5 \}$ and 
\be \label{qy_cl}
X'_L=X'_R={\rm Diag.}\left(0,1,-1,1,-1\right),\,  Y'_L={\rm
Diag.}\left(-\frac{1}{2},-\frac{1}{2},-\frac{1}{2},-1,-1\right),\, 
Y'_R=-{\bf 1}\,.\ee
Using Eq. (\ref{XY}) and definition in Eq. (\ref{UVdef}), the gauge
couplings in the mass basis are obtained as
\be \label{YLR_cl}
Y_L = -\frac{1}{2} U_L^\dagger U_L - V_L^\dagger V_L\,,\,\, Y_R = -{\bf
1}\,,\ee
\be \label{QLR_cl}
X_{L,R} = U_{L,R}^\dagger\, X_3\, U_{L,R} + V_{L,R}^\dagger\, X_2\,
V_{L,R}\,,\ee
where $X_2 = {\rm Diag.}(1,-1)$. Eq. (\ref{qy_cl}) implies ${\rm Tr}(Y'_L X'_L + Y'_R X'_R)=0$ making $A_{BZ'}$ finite in the present framework. Moreover, using Eqs.
(\ref{YLR_cl},\ref{QLR_cl}) and the relations obtained in Eqs.
(\ref{rel1_cl},\ref{rel2_cl}) we find 
\beqa\label{}
& & \frac{1}{2} {\rm Tr}(m^2(\{Y_L,X_L\}+\{Y_R,X_R\})) - {\rm Tr}(Y_L m X_R
m + Y_R m X_L m)  \nonumber \\
& & = \frac{1}{2}{\rm Tr}\left( (M_l M_l^\dagger  - M_l^\dagger M_l  +m_l m_l^\dagger) X_3 - m_l^\dagger m_l X_2  \right)\,,
\eeqa
which vanishes identically for the above considered forms of $M_l$ and $m_l$. Therefore, the divergent part in $B_{BZ'}$ also vanishes making the KM finite and calculable in the underlying framework. The values of $A_{BZ'}$ and $B_{BZ'}$ can be explicitly computed using the expressions of finite parts given in Eqs. (\ref{A_gen}, \ref{B_gen}) with $Y_{L,R}$ and $X_{L,R}$ obtained in Eqs.  (\ref{YLR_cl},\ref{QLR_cl}) for this model.

We explicitly calculate the KM in a specific ``seesaw-like"
case, i.e. $M_l < m_l, \tilde{m}_l \ll M_f$, within this model. The
effective mass matrix for the three light charged leptons is obtained as
${M_l}^{\rm eff.} \approx M_l - m_l M_f^{-1} \tilde{m}_l$. Let $u_{L,R}$ and
$v_{L,R}$ are matrices which diagonalize ${M_l}^{\rm eff.}$ and $M_f$,
respectively, such that 
\be \label{}
u_L^\dagger\, {M_l}^{\rm eff.}\, u_R = {\rm
Diag.}(m_e,m_\mu,m_\tau)\,,~~v_L^\dagger\, M_f\, v_R = {\rm Diag.}(m_4,
m_5)\,. \ee
The $5 \times 5$ unitary matrices ${\cal U}_{L,R}$ can suitably written as
\be \label{}
{\cal U}_{L,R} \approx \left( \ba{cc} u_{L,R} & -\rho_{L,R}\, v_{L,R} \\
\rho_{L,R}^\dagger\, u_{L,R} &  v_{L,R} \ea\right) \,, \ee
where $\rho_L \approx -m_l M_f^{-1}$ and $\rho_R^\dagger \approx - M_f^{-1}
\tilde{m}_l$. 
Further, we take $\lambda^l_{e\mu}=\lambda^l_{\mu e} =0$ and consider the
following ansatz for the matrices $m_l$ and $\tilde{m}_l$:
\be 
m_l = m\,\left(\ba{cc} 0 & 0 \\ 1 & 0 \\ 0 & 1 \ea \right)\,, \,\,
\tilde{m}_l = \tilde{m}\, \left(\ba{ccc} 0 & 1 & 0 \\ 0 & 0 & 1 \ea
\right)\,.
\ee
The above forms are achieved if \lmt symmetry remains unbroken in $m_l$, $\tilde{m}_l$. In this case, the unitary matrices $u_{L,R}$ and $v_{L,R}$ can be
parametrized as
\be \label{}
u_{L,R} = \left( \ba{ccc} 1 & 0 & 0\\ 0 & c_{\theta_{L,R}} &
s_{\theta_{L,R}} \\ 0 & -s_{\theta_{L,R}}& c_{\theta_{L,R}}
\ea\right)\,,~~v_{L,R} = \left( \ba{cc}  c_{\phi_{L,R}} & s_{\phi_{L,R}} \\ 
-s_{\phi_{L,R}} & c_{\phi_{L,R}} \ea\right)\,,\ee
where $c_\theta = \cos\theta$ and $s_\theta = \sin\theta$. The general results given in Eqs. (\ref{A_gen},\ref{B_gen}) are then used to compute the KM using the above simplifications. The leading contributions to kinetic and mass mixing are obtained as:
\beqa \label{AB_cl}
A_{BZ'} &\approx & -\frac{g_Yg'}{4 \pi^2}\Bigg((c_{2
\phi_L}+c_{2\phi_R})\left(b_2[m_4,m_4,q] -  b_2[m_5,m_5,q] \right) \Bigg.
\nonumber \\
&+& \Bigg.. \frac{1}{2}(c_{2 \theta_L}+2 c_{2\theta_R})
\left(b_2[m_\mu,m_\mu,q] -  b_2[m_\tau,m_\tau,q]\right) +  {\cal O}\left(
\frac{m^2,\tilde{m}^2}{m_{4,5}^2}\right)\Bigg)\,, \nonumber \\
B_{BZ'} &\approx & \frac{g_Yg'}{16
\pi^2}\Bigg(\left(m^2-2\tilde{m}^2\right)(c_{2
\phi_L}-c_{2\phi_R})(b_0[m_4,m_4,q] - b_0[m_5,m_5,q]) \Bigg. \nonumber \\
&+& \Bigg. (c_{2 \theta_L}- c_{2\theta_R}) \left(m_\tau^2\,
b_0[m_\tau,m_\tau,q]-m_\mu^2\, b_0[m_\mu,m_\mu,q]\right) +  {\cal O}\left(
\frac{m^2,\tilde{m}^2}{m_{4,5}^2}\right)\Bigg)\,. \eeqa

For $|q^2| \ll m_5^2, m_4^2$, the above expressions can further be
simplified as
\beqa \label{ABsimp_cl}
A_{BZ'} &\approx & -\frac{g_Yg'}{24 \pi^2}\Bigg((c_{2
\phi_L}+c_{2\phi_R})\,\ln \frac{m_4^2}{m_5^2} + \frac{1}{2}(c_{2 \theta_L}+2
c_{2\theta_R})\, \ln \frac{m_\mu^2}{m_\tau^2}\Bigg)\,, \nonumber \\
B_{BZ'} &\approx & \frac{g_Yg'}{16
\pi^2}\Bigg(\left(m^2-2\tilde{m}^2\right)(c_{2 \phi_L}-c_{2\phi_R})\ln
\frac{m_4^2}{m_5^2} \Bigg. \nonumber \\
&+& \Bigg. (c_{2 \theta_L}- c_{2\theta_R}) \left(m_\tau^2\,
\ln\frac{m_\tau^2}{\mu^2}-m_\mu^2\, \ln\frac{m_\mu^2}{\mu^2}\right)\Bigg)\,.
\eeqa
The first terms in $A_{BZ'}$ and $B_{BZ'}$ in Eq. (\ref{ABsimp_cl}) quantify
the 1-loop contribution induced by the vectorlike charged leptons. Since
these fermions are charged under both the $U(1)_Y$ and \lmt gauge
symmetries, their contribution to KM is nonzero unless $M_f$ is \mt
symmetric, i.e. $\phi_{L,R}=\pi/4$ or $m_4=m_5$. This is in contrast to the
standard seesaw case discussed in the previous subsection where the RH
neutrinos do not couple to $B$ and hence they do not induce KM by
themselves. The second terms in $A_{BZ'}$ and $B_{BZ'}$ correspond to
contributions from the SM charged leptons. This along with the other
sub-leading contributions in Eq. (\ref{AB_cl}) vanish in the decoupling
limit, $m_{4,5} \to \infty$. Note that  ${M_l}^{\rm eff.} = M_l$ is \mt
symmetric in the same limit which leads to $\theta_{L,R} = \pi/4$ and
vanishing of the charged lepton contributions. In this case, the first terms
in $A_{BZ'}$ and $B_{BZ'}$ provide non-decoupling contributions to the
kinetic and mass mixing respectively.

$M_f$ is a diagonal matrix in the \lmt symmetric limit. As a result, the masses of vectorlike leptons need not be linked to the \lmt symmetry breaking scale, unlike in the seesaw case discussed in the previous subsection. This allows the \lmt breaking scale to be smaller than the vectorlike lepton masses which are required to be large for the phenomenological reasons. The $Z^\prime$ boson in this case can be light and leave signal in the low energy process. The other advantage of this is that one gets almost diagonal $M_f$ resulting in $\phi_{L,R} \approx 0$. If one also assumes that the elements of $M_l$ are vanishingly small and the second and the third generation masses arise through the seesaw like contribution 
\be \label{}
{M_l}^{\rm eff.} \approx - m_l M_f^{-1} \tilde{m}_l \approx - m \tilde{m}\, 
{\rm Diag.}(m_4^{-1}, m_5^{-1})\, \ee
then this leads to seesaw contribution which is almost diagonal and results in small $\theta_{L,R}$. With a small $\theta_L$ the atmospheric mixing gets dominant  contribution from the \mt symmetric
neutrino mass matrix and is nearly maximal as required phenomenologically. Further, $m_4/m_5 \approx m_\tau/m_\mu$ if $\phi_L$ is vanishingly small. Replacing these in Eq. (\ref{ABsimp_cl}) results in
\be \label{}
A_{BZ'} \approx \frac{g_Yg'}{48 \pi^2} \ln
\frac{m_\mu^2}{m_\tau^2}\,,~~B_{BZ'} \approx {\cal
O}\left(\frac{m^4,\tilde{m}^4}{m_{4,5}^2}\right)\,.\ee
The \mt symmetry in the charged lepton sector is badly broken giving rise to
large but finite $A_{BZ'}$. The leading order contribution to mass mixing
parameter $B_{BZ'}$ vanish in this case because of $\phi_L \approx \phi_R$.

The above setup can straightforwardly be implemented in the quark sector
extending \lmt symmetry to include the second and third generations of
quarks transforming in an analogous way. The \mt interchange symmetry can
also be generalized as 2-3 interchange symmetry \cite{Joshipura:2005vy}. The
up-type quark mass matrix can be assumed invariant under 2-3 interchange
symmetry. A similar assumption for the down-type quarks would then imply
$V_{cb}=V_{ub}=0$ and therefore breaking of 2-3 interchange symmetry would be
necessarily required for the realistic quark mixing angles. Such breaking can be
incorporated by extending the down-type quark sector by a pair of vectorlike
quarks in an analogous way discussed above. One obtains similar expressions
for $A_{BZ'}$ and $B_{BZ'}$ as in Eq. (\ref{ABsimp_cl}) with appropriate
change in hypercharges and an overall color factor. The difference compared
to the leptonic case is that one requires a small deviation from
$\theta_{L}=\pi/4$ in order to produce realistic quark mixing. This can easily be reproduced through small seesaw-like contribution, and one need not assume
vanishing $M_l$ as it is done in the leptonic case. The mild breaking of 2-3
interchange symmetry gives rise to relatively small kinetic and mass mixing in this case.

\section{Summary and Discussions}
\label{sec:summary}
The SM extended with gauged \lmt symmetry offers phenomenologically rich
framework in which the new physics effects can arise directly through the
couplings of the second and the third generation leptons with the $Z'$ boson
and indirectly through the KM between the $Z$ and $Z'$ bosons.
The later makes it possible for the SM quarks and the first generation of leptons to couple to $Z'$ boson and therefore the KM is of particular interests from the phenomenological considerations. KM in the standard \lmt models can be forbidden to all orders if one imposes \mt interchange or
reflection symmetry under which one of the two gauge bosons also transforms
non-trivially. However, the same symmetries do not give phenomenologically
viable mixing in the lepton sector. Invariance of leptonic Lagrangian under
\mt interchange symmetry leads to vanishing atmospheric and reactor mixing
angles while the same under \mt reflection symmetry implies CP conservation
in the lepton sector. We showed that it is possible to create more general versions of these symmetries, which can lead to realistic lepton mixing. However, these symmetries can forbid the KM up to 1-loop level only.

In the absence of \mt symmetry, the kinetic and mass mixing in \lmt models is given by arbitrary parameters which cannot
be determined from the other fundamental parameters of the theory and can be constrained only from the experimental observations. However, the KM parameters become calculable if \mt symmetry is imposed in a full theory, and the mechanism of its breaking is known. The magnitude of KM in this case depends on the details of the new sector responsible for \mt breaking. We provided two explicit examples of this in section \ref{sec:model}. Both the kinetic mixing parameters are shown to be small and inversely proportional to the right handed neutrino masses  in a class of seesaw models in which  the \mt symmetry breaking is present only in the heavy neutrino sector. The neutrino mass mixing parameter can be large and independent of the right handed neutrino masses if the Dirac mass matrix also break the \mt symmetry. On the contrary if \mt breaking is introduced through heavy vectorlike charged leptons, the KM parameters are dominantly determined by the new sector, and its magnitude can be large.

Phenomenological consequences of the SM with \lmt extensions are widely discussed and used to constrain the parameters $g'$ and $M_{Z'}$. 
One could divide the tests of this model in two categories, one which exploits KM and use electron or hadron induced interactions. 
These include a large variety of processes, such as precision electroweak tests, atomic parity violation, beam dump experiments, 
$\nu_e-e$ elastic scattering in Borexino, and coherent neutrino-nucleus scattering observed in the COHERENT experiment.  
Constraints from these experiments mainly for light $Z'$ are presented in \cite{Bauer:2018onh}. These constraints do not hold in the
type of seesaw model discussed in section \ref{sec:model_I} due to very suppressed $Z$-$Z'$ mixing. 
The other class of tests involve only $\mu$ and $\tau$ sector. Anomalous magnetic moment of $\mu$ and $\tau$ and the muon 
neutrino indued trident production through the process $\nu N\rightarrow \nu N \mu^+\mu^-$ fall in this category, and have been used to constrain the purely leptonic couplings of $Z'$. The latter process is found to be quite constraining and rules out most of the parameter space corresponding to $M_{Z'}>400$ MeV which otherwise can be used to explain the discrepancy in $(g-2)_\mu$. It turns out that the neutrino trident process is not a useful probe of models considered here and in \cite{Foot:1994vd,Baek:2001kca}. The general \mt reflection symmetry requirement in Eq. (\ref{mutaurgen}) imposed to get vanishing KM parameters at 1-loop also implies that the $Z'$ couplings to leptons are purely off-diagonal in their mass basis. Immediate consequence is that the trident process $\nu N\rightarrow \nu N \mu^+\mu^-$ cannot take place at tree level and is unable to constrain the parameters of the model. Instead, the rare tau decays could provide stringent constraint on the
model. If \lmt symmetry is broken only through Higgs doublet VEV then 
the rare decay $\tau\rightarrow\mu Z'$ (for light $Z'$) and the decay $\tau ^-\rightarrow  \mu^- \overline{\nu}_\mu \nu_\tau$ 
together rule out the entire space which is responsible for the explanation of $(g-2)_\mu$  \cite{Foot:1994vd,Baek:2001kca}. 
Small parameter space is still allowed if $SU(2)$ singlet field is responsible for the \lmt breaking as is assumed here. 
This is analyzed in \cite{Baek:2001kca}. The flavour off-diagonal couplings of $Z^\prime$ to the charged leptons,
as obtained in Eq. (\ref{falpha}), depend on the exact  structure of the mixing matrices $U_{l_L}$ and $U_{l_R}.$ These matrices have the form given in Eq. (\ref{umt}) in the limit of the exact \mt symmetry. This gives the following couplings of $Z^\prime$:
\be \label{zpch}
 g^\prime Z_\mu^\prime( \cos\theta_L\, \overline{\tau}_L\gamma^\mu \mu_L+ \sin\theta_L\, \overline{\tau}_L\gamma^\mu e_L+\cos\theta_R\, \overline{\tau}_R\gamma^\mu \mu_R+ \sin\theta_R\, \overline{\tau}_R\gamma^\mu e_R)\,, \ee
where $\theta_{L,R}$ are angles entering in definitions of $U_{lL,lR}$ as given in Eq. (\ref{umt}). This equation implies non-universality in the decay of $\tau$ to $e$ and $\mu$. The above equation coincides with the one assumed in \cite{Foot:1994vd,Baek:2001kca} for $\theta_L=\theta_R=0.$ It is found in \cite{Baek:2001kca} that the above coupling can explain the $(g-2)_\mu$ anomaly and be consistent with the observed rare tau decay $ \tau ^-\rightarrow  \mu^- \overline{\nu}_\mu \nu_\tau$  for a very narrow ranges in parameters $g^\prime$ and $M_{Z^\prime}$. We update their analysis considering the latest values of  $(g-2)_\mu$ from \cite{Broggio:2014mna} and BR($\tau ^-\rightarrow  \mu^- \overline{\nu}_\mu \nu_\tau$) from \cite{Altmannshofer:2014cfa}. We observe that the positive deviation at $1.6\sigma$  found in the branching ratio of the decay  $\tau ^-\rightarrow  \mu^- \overline{\nu}_\mu \nu_\tau$  compared to its SM value and the anomaly in $(g-2)_\mu$ can be simultaneously reconciled for 
$$ 0.004 \le g^\prime \le 0.006 ~~{\rm and }~~1.12\, {\rm GeV} \le M_{Z^\prime} \le 1.24\, {\rm GeV}$$
which practically coincides with the one already found in \cite{Baek:2001kca}. 
One can obtain $M_{Z^\prime}\sim  g^\prime v_s\sim {\cal O}(1)~{\rm GeV}$ for the \lmt breaking scale around TeV if the above range
in parameters is to be realized.  Possible constraint on this model can come at the muon
collider \cite{Baek:2001kca} through the process $\mu^+\mu^-\rightarrow
\tau^+\tau^-$, search for rare tau decays at Belle II and detection of four
charged leptons at colliders as discussed in details in \cite{Altmannshofer:2014cfa}.

\acknowledgments
The work of KMP was partially supported by research grant under INSPIRE Faculty Award (DST/INSPIRE/04/2015/000508) from the Department of Science and Technology, Government of India.
 
\begin{appendix}
\section{Loop integrals}
\label{sec_app}
The definition of the loop integration functions are as the following \cite{Peskin:1995ev}.
\beqa \label{app_int}
b_0[m_i,m_j,q] &=& \int_0^1\, dx\, \ln(\Delta(m_i,m_j,q)/\mu^2)\,, \nonumber \\
b_1[m_i,m_j,q] &=& \int_0^1\, dx\, x\, \ln(\Delta(m_i,m_j,q)/\mu^2)\,, \nonumber \\
b_2[m_i,m_j,q] &=& \int_0^1\, dx\, x(1-x)\,\ln(\Delta(m_i,m_j,q)/\mu^2)\,,\eeqa
where $\Delta(m_i,m_j,q) = x m_j^2+(1-x) m_i^2 - x(1-x) q^2$. $\mu$ is an arbitrary subtraction scale. The functions $b_0$ and $b_2$ are symmetric under the interchange of $i$ and $j$. Also, $b_1[m_i,m_i,q] =\frac{1}{2}b_0[m_i,m_i,q]$. 

For special cases of interests, the above integrals can be approximated as the following.
For $m^2 \ll q^2$,
\beqa\label{app_int_1}
 b_0[m,m,q] &\approx & - 2 + \ln\left(-\frac{q^2}{\mu^2}\right)-\frac{m^2}{q^2}\, \nonumber \\
b_2[m,m,q] & \approx& -\frac{5}{18}+\frac{1}{6} \ln\left(-\frac{q^2}{\mu^2}\right)-\frac{m^2}{q^2}\,.
\eeqa

For $m^2 \ll q^2 \ll M^2$ (${\cal{O}}(m^2/q^2)$ terms not shown),
\beqa\label{app_int_2}
b_0[m,M,q] &\approx & -1+\ln \frac{M^2}{\mu^2} -\frac{q^2}{2 M^2}\,,\nonumber \\
b_1[m,M,q] &\approx & -\frac{1}{4} + \frac{1}{2} \ln\frac{M^2}{\mu ^2} -\frac{q^2}{6 M^2}\,,\nonumber \\
b_1[M,m,q] &\approx & -\frac{3}{4} + \frac{1}{2} \ln\frac{M^2}{\mu ^2} -\frac{q^2}{3 M^2}\,,\nonumber \\
b_2[m,M,q] &\approx & -\frac{5}{36} + \frac{1}{6} \ln\frac{M^2}{\mu ^2}-\frac{q^2}{12 M^2}\,.
\eeqa

For $q^2 \ll M^2$,
\beqa\label{app_int_3}
b_0[M,M,q] &\approx & \ln\frac{M^2}{\mu ^2}-\frac{q^2}{6 M^2}\,,\nonumber \\
b_1[M,M,q] &\approx & \frac{1}{2}\ln\frac{M^2}{\mu ^2}-\frac{q^2}{12 M^2}\,,\nonumber \\
b_2[M,M,q] &\approx & \frac{1}{6}\ln\frac{M^2}{\mu ^2}-\frac{q^2}{30 M^2}\,.
\eeqa
\end{appendix}

\bibliography{refs.bib}
\bibliographystyle{apsrev4-1}
\end{document}